\documentclass[10pt]{article}
\usepackage{graphicx}
\usepackage{amsmath}
\usepackage{amssymb}
\usepackage{epstopdf}

\usepackage{cite}

\usepackage{hyperref}

\usepackage{lineno}

\usepackage{microtype}
\DisableLigatures[f]{encoding = *, family = * }

\usepackage[labelfont=bf,labelsep=period,justification=raggedright]{caption}

\usepackage{color}

\usepackage{siunitx}

\makeatletter
\renewcommand{\@biblabel}[1]{\quad#1.}
\makeatother

\date{}

\begin{document}

\begin{flushleft}
{\Large
\textbf{Mean Field Approach for Configuring Population Dynamics on a Biohybrid Neuromorphic System}
}
\\
Johannes Partzsch$^{1,\ast,3}$, 
Christian Mayr$^{1,3}$,
Massimiliano Giulioni$^{2}$, 
Marko Noack$^{1}$,
Stefan H\"anzsche$^{1}$,
Stefan Scholze$^{1}$,
Sebastian H\"oppner$^{1}$,
Paolo Del Giudice$^{2}$,
Rene Sch\"uffny$^{1}$
\\
\bf{1} Chair for Highly Parallel VLSI Systems and Neuromorphic Circuits, Department of Electrical Engineering and Information Technology, Technische Universit\"at Dresden, Dresden, Germany
\\
\bf{2} Department of Technologies and Health, Istituto Superiore di Sanita, Roma, Italy
\\
$\ast$ E-mail: Johannes.Partzsch@tu-dresden.de
\\
\bf{3} These two authors contributed equally
\end{flushleft}

\section*{Abstract}

Real-time coupling of cell cultures to neuromorphic circuits necessitates a neuromorphic network that replicates biological behaviour both on a per-neuron and on a population basis, with a network size comparable to the culture. We present a large neuromorphic system composed of 9 chips, with overall 2880 neurons and 144M conductance-based synapses. As they are realized in a robust switched-capacitor fashion, individual neurons and synapses can be configured to replicate with high fidelity a wide range of biologically realistic behaviour. In contrast to other exploration/heuristics-based approaches, we employ a theory-guided mesoscopic approach to configure the overall network to a range of bursting behaviours, thus replicating the statistics of our targeted in-vitro network. The mesoscopic approach has implications beyond our proposed biohybrid, as it allows a targeted exploration of the behavioural space, which is a non-trivial task especially in large, recurrent networks.


\section*{Introduction}

Real-time neuromorphic systems allow for a direct coupling with biological tissue \cite{mazurek12,serb2017}, enabling to understand, gently control and virtually extend the biological part. Seamless dynamical integration of hardware and biology makes such a hybrid system most effective, where we define seamless as that the hardware neural network operates in the same dynamical regime as its biological counterpart, and tight coupling of both generates a meaningful joint dynamics.

As a prerequisite, the neuromorphic hardware should allow to implement finely tunable dynamical modes comparable to biology, for example exhibiting asynchronous firing and being able to generate network bursts \cite{gigante15,levi12}. A wide range of theoretical works exists on how to generate these dynamical regimes \cite{brunel00,renart03,gigante15}. However, this necessitates both a neuromorphic network that is reasonably close to a given theoretical model and a method for tuning its behaviour to a desired regime. Most neuromorphic circuits so far have used sub-threshold circuits \cite{indiveri10, park14}, which exhibit high sensitivity to device mismatch, process variation and temperature, and therefore are hard to control \cite{henker07}. As an alternative, OTA-based solutions have been proposed \cite{noack10,koickal07}, which, however, struggle with biological real-time capability in large-scale integrated systems where only minimum area is available \cite{mayr2014ota}.

As an alternative to the above, we have designed a neuromorphic system based on switched-capacitor (SC) circuits. In SC circuits, time constants and gain parameters depend on capacitance ratios and switching frequencies and not on process-dependent transistor parameters. Capacitance ratios can be manufactured with high precision \cite{allen02}, and the switching frequency can be controlled and finely tuned by digital circuits, allowing for faithful reproduction of model parameters and successful implementations in modern process technologies despite increased device mismatch \cite{mayr15,noack15}. The system allows replication of realistic conductance synapses (e.g. NMDA, GABA, AMPA) and spike frequency adaptation 
as well as Markram/Tsodyks type presynaptic adaptation 
As we are only interested in short-term dynamics comparable to our in-vitro network, we have omitted long-term plasticity. 

With this system, we employ a theory-guided, mean-field approach to predict recurrent behaviour based on open-loop characterization of the neuromorphic network. We show which parameters of the open loop transfer function govern which behavioural aspects of the recurrent network, thus enabling a detailed steering of the targeted behaviour, with very close agreement between theory and neuromorphic hardware. 


We demonstrate the capability of this system to faithfully reproduce theoretical results, 
to show a wide range of dynamical regimes in hardware, especially concentrating on bursting behavior as seen in cultured networks \cite{levi12}.
without need for time-consuming individual parameter calibration. 
The latter also saves hardware resources for single-neuron configuration and calibration, e.g. reducing storage space for individual parameters.

In the following, we first introduce the system architecture and the employed circuits, motivating design choices from general assumptions of mean-field theory.
We then describe the employed mean-field approach and analysis methods for the expected dynamical regimes. In Results, we then show measurements of transfer curves for characterizing general network behavior, and then move on to a systematic parameter space exploration, showing different bursting regimes.



\section*{Materials and Methods}

\subsection*{Chip Architecture and Circuit Components}

In mean-field theory, neurons are treated as being statistically equivalent.
In consequence, all neurons of a population typically have the same base parameters, like synaptic and membrane time constants or firing threshold, allowing to share them between neurons.
This property is key for utilizing the multi-synapse approach \cite{vogelstein07,benjamin14}, where one synapse circuit represents a set of synapses with the same properties, being driven by their joint input spiking activity.
This approach reduces total silicon area significantly, because the number of synapse circuits is drastically reduced.
Furthermore, it is more flexible, because there are no hard bounds on the number of synapses per neuron, in contrast e.g. to synapse matrix architectures \cite{partzsch15}.

\begin{figure}
 \centering
 \includegraphics[scale=0.35]{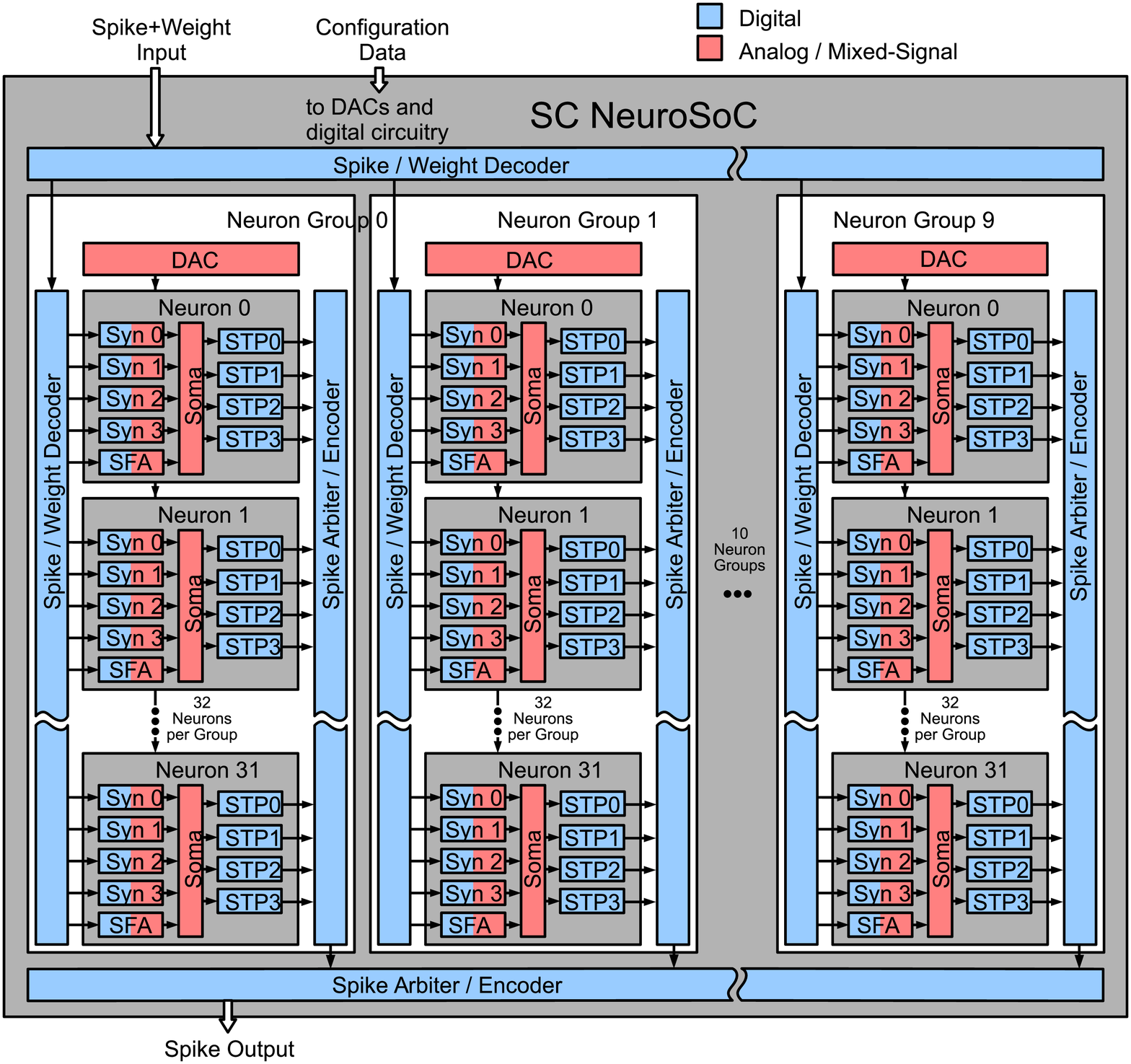}
 \caption{\label{fig_chiparch} Chip architecture with multi-synapses}
\end{figure}

Fig. \ref{fig_chiparch} shows the architecture of the SC NeuroSoC, which follows the multi-synapse approach.
It comprises 10 neuron groups each with 32 neuron circuits.
All neurons of one group share the same set of parameters, saving significantly on silicon area for parameter storage.
Spike decoding and arbitration is done in a hierarchical manner. First, an incoming pulse packet is routed to the appropriate neuron group and then to the targeted synapse of one neuron.
Once a neuron produces a spike, it is forwarded to four digital short-term plasticity (STP) circuits, implementing the quantal release model \cite{markram98}.
Placing the STP circuit at the output of the neuron saves silicon area overall, because the STP output is calculated only once per source neuron in the system.
Using four STP circuits allows four different parameter sets to be used, which offers enough flexibility for most practical applications.
Each STP circuit produces a 6~bit output weight, which is forwarded together with the neuron's address to the spike output of the chip.
For the SC circuits, each neuron group is equipped with a digital-to-analog converter, which provides the reversal potentials and firing threshold voltages, equal for all neurons in the group.

\begin{figure}
 \centering
 \includegraphics[scale=0.45]{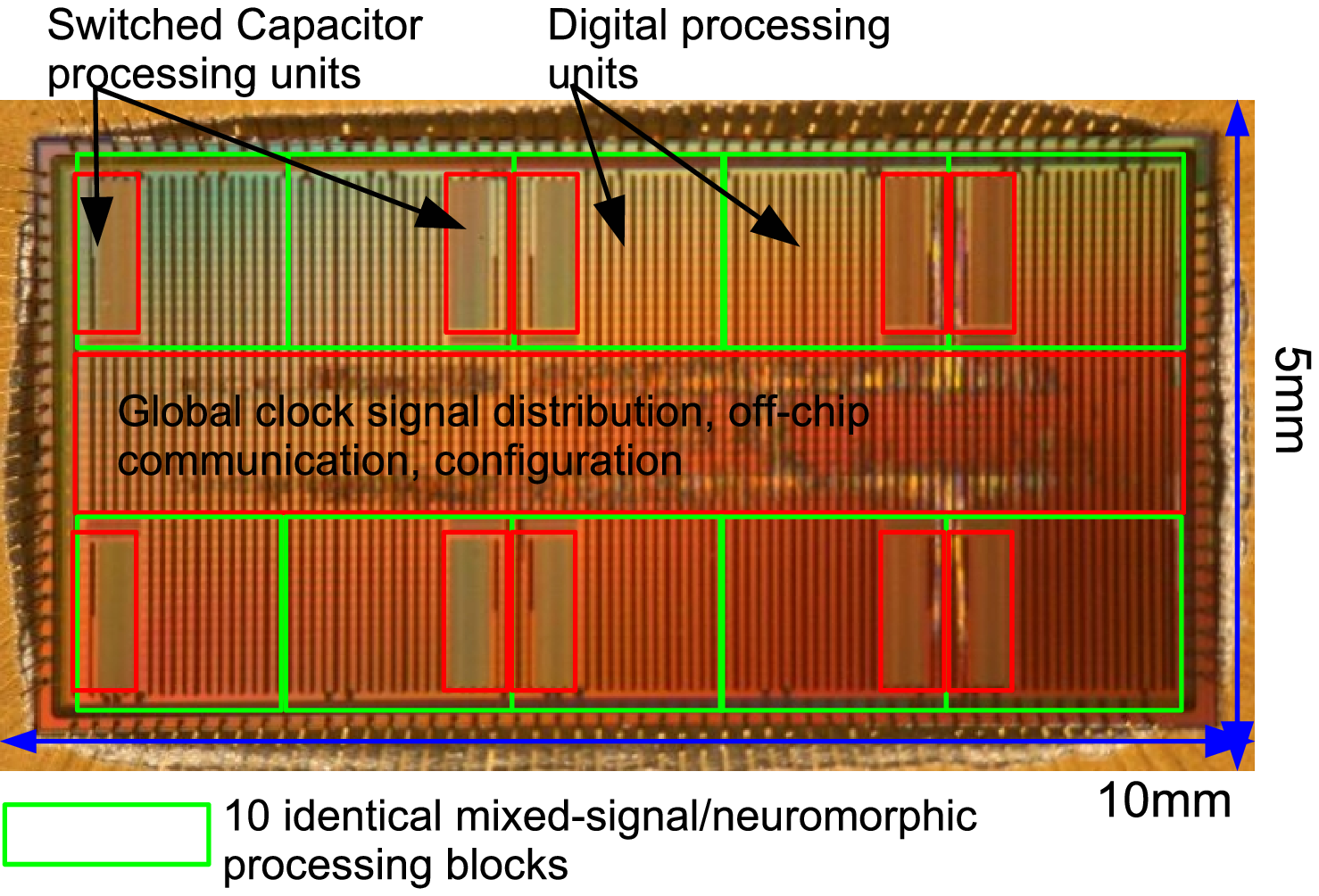}
 \caption{\label{fig_bionect_die} Chip microphotograph}
\end{figure}

Fig. \ref{fig_bionect_die} shows a chip photograph with annotated switched-capacitor processing units, digital processing units and global clock signal distribution, off-chip communication and configuration circuitry. The global clock is supplied externally to reduce chip complexity and allow greater configurability compared to an on-chip frequency generator \cite{eisenreich09}.
The chip was implemented in a UMC 180nm technology.
Its size is $\SI{10}{\milli\meter} \times \SI{5}{\milli\meter}$.

\begin{figure}
\centering
\includegraphics[scale=0.6]{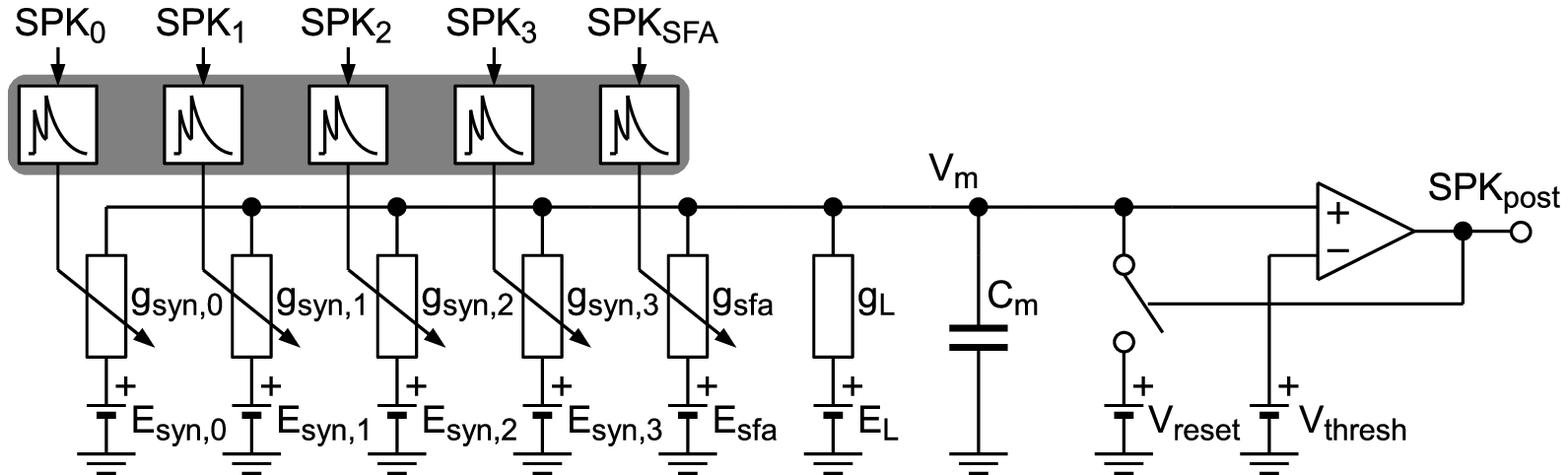}
\caption{Leaky integrate-and-fire neuron with different types of conductance-based multi-synapses and spike-frequency adaptation.}
\label{fig:circuit_overview}
\end{figure}

A simplified schematic of the neuron circuit is depicted in Fig. \ref{fig:circuit_overview}.
It shows a leaky integrate-and-fire neuron with four individually configurable conductance-based multi-synapses.
A fifth multi-synapse is added for implementing spike-frequency adaptation (SFA).
Its circuit is identical to the other multi-synapses, so that it can be used to model a fifth synapse type if no SFA is required.
Additionally, one of the four multi-synapses is extended by a voltage-dependent reversal potential, modeling the behavior of NMDA-type synapses.
Details of this circuit can be found in \cite{noack12}.

\begin{figure}
\centering
\includegraphics[scale=0.45]{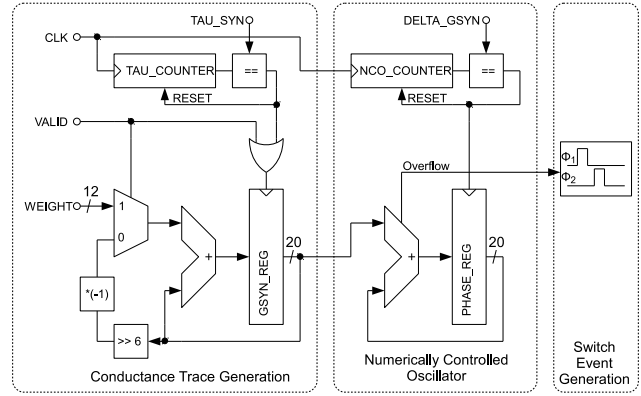}
\caption{Block diagram of the digital circuitry.}
\label{fig:digital}
\end{figure}

One multi-synapse can represent a large amount of individual synapses which share the same reversal potential and time constant of the synaptic conductance.
The conductance trace can be modeled by instantaneous jumps of a certain height at incoming pulses and an exponential decay between spikes \cite{rolls10}.
This behavior is modeled by a digital circuit in our implementation.
It is shown in the left part of Fig. \ref{fig:digital}.
At an incoming pulse, VALID goes high and an associated 12 bit weight is provided, which is accumulated in the register GSYN\_REG. The clock signal runs continuously and lets TAU\_COUNTER count upwards until its value is equal to TAU\_SYN.
Then the counter is reset and GSYN\_REG is attenuated by a factor of $(1-2^{-6})\approx 0.984$, which is done by a right shift operation and a subtraction. This results in an exponential decay with a time constant depending on TAU\_SYN and the global clock frequency.

The synaptic conductance itself is realized via an SC circuit.
Its conductance is given by $g_\mathrm{syn}=C_\mathrm{syn}\cdot f$, where $C_\mathrm{syn}$ is the switching capacitance in the synapse circuit and $f$ is the switching frequency.
Thus, the conductance value in register GSYN\_REG needs to be converted to a switching frequency $f$.
This is done by the numerically controlled oscillator shown in the right part of Fig. \ref{fig:digital}.
The conductance value GSYN\_REG is accumulated in PHASE\_REG with a period defined by DELTA\_GSYN, which controls the conductance scaling.
When an overflow occurs, a switch event for the SC circuit is generated, triggering two non-overlapping switch signals $Phi_1$ and $Phi_2$.
Via this simple circuit, the switching frequency $f$ follows the value of GSYN\_REG proportionally.

The digital circuit shown in Fig. \ref{fig:digital} thus guarantees that an incoming spike train is translated into a biologically realistic conductance trace and from there to a series of switch events for the respective synapse SC circuit.
This is in contrast to the system introduced in \cite{vogelstein07}, where each input spike to the system directly triggers a switch event for the SC circuit.
This would mean that realistic conductance traces had to be generated off-chip, resulting in a multiplication of the input spike rate.

\begin{figure}
\centering
\includegraphics[scale=0.4]{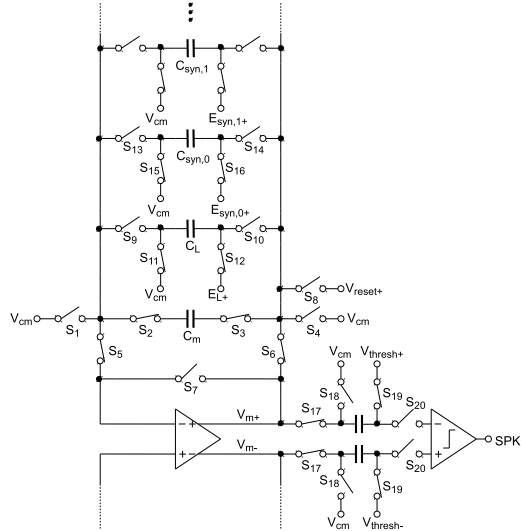}
\caption{SC neuron circuit with conductance-based synapses and comparator for threshold detection.}
\label{fig:circuit}
\end{figure}

In Fig. \ref{fig:circuit} the analog circuitry can be seen, consisting of the membrane capacitor $C_m$ and 5 capacitors $C_{syn,1-5}$, which emulate the synaptic conductance of the different multi-synapse types. $C_{syn,1}$ is surrounded by the switches $S_{15}$ and $S_{16}$, which are closed at $Phi_1$ and $S_{13}$ and $S_{14}$, which are closed at $Phi_2$ according to the non-overlapping switch signals generated by the digital circuitry as shown in Fig. \ref{fig:digital}. At $Phi_1$ $C_{syn,1}$ is charged by the corresponding reversal potential $E_{syn,1+}$ and at $Phi_2$ a charge equalization between $C_{syn,1}$ and $C_m$ is performed, which lets the membrane potential decay towards the reversal potential. The other synapses work analogously. $C_L$ models the leakage of the membrane and therefore is also switched in a similar way as the synapses, but with a constant switching frequency.
The additional switches $S_{1}$ and $S_{4}$ have been introduced to reduce leakage between switching events \cite{noack12,noack15,mayr15}.

In contrast to \cite{vogelstein07}, all capacitors are instantiated twice, because the circuit comprises a fully differential design which reduces charge injection and clock feed-through and doubles the usable voltage range. The differential membrane potential is buffered by an operational amplifier which allows monitoring every neuron on the chip with an oscilloscope.
Moreover, the buffered membrane voltage is used for implementation of the NMDA voltage dependence in one of the multi-synapses, as described in \cite{noack12}.

\subsection*{System Integration}

\begin{figure}
 \centering
 \includegraphics[scale=0.45]{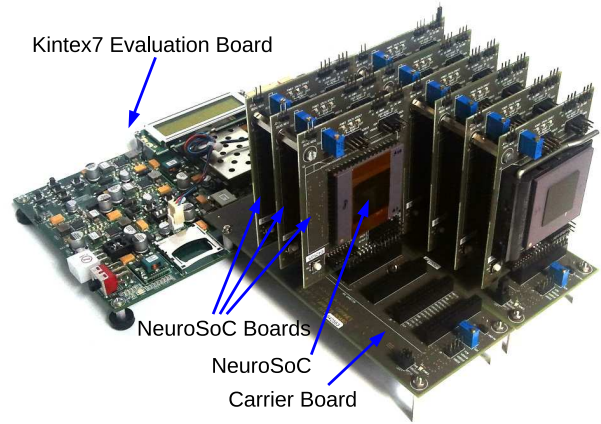}
 \caption{\label{fig_setup} Photograph of the system setup}
\end{figure}


With its dedicated pulse input and output interfaces, the NeuroSoC is designed for operating together with a field-programmable gate array (FPGA), which can be used to connect several NeuroSoCs.
For this, a Xilinx KC705 evaluation board with a Xilinx Kintex7 FPGA has been extended with custom printed circuit boards.
A carrier board connects to one of the extension headers of the FPGA evaluation board, generating supply voltages for the NeuroSoCs and distributing signals to six smaller extension headers.
On each of these headers, one NeuroSoC board can be plugged in, holding a socket for one NeuroSoC and providing pin headers for debug outputs.
The FPGA evaluation board features one high pin count and one low pin count extension header, the latter only providing IO pins for three NeuroSoCs.
Thus, a total of nine NeuroSoCs may be connected to one FPGA, forming a system with 2880 neurons.
Figure \ref{fig_setup} shows a photograph of the complete setup.

The Kintex7 FPGA is the main hub in the system, connecting the NeuroSoCs among each other.
It provides a Gbit-Ethernet link for interfacing to a host PC for configuration, and for communicating with other spiking systems via the user datagram protocol (UDP), such as other neuromorphic systems, real-time software pulse generators, or micro-electrode arrays for interfacing to biological tissue.
For this, two previously developed protocols for pulse exchange \cite{rast13,george15} are supported.
Furthermore, the FPGA contains buffers for pulse stimulation and tracing to/from single NeuroSoCs, which are interfaced via Gbit-Ethernet as well.

\begin{figure}
 \centering
 \includegraphics[scale=0.45]{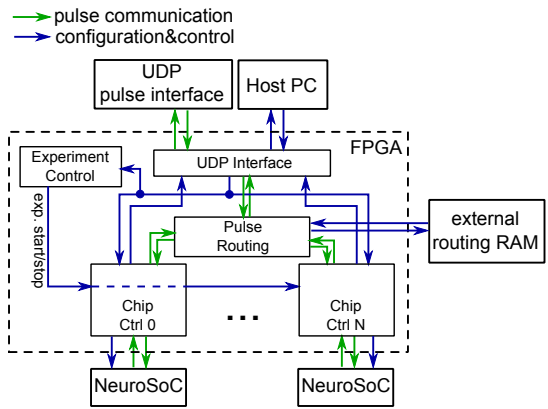}
 \caption{\label{fig_sysarch} System architecture and main FPGA components}
\end{figure}

The structure of the FPGA firmware and its connections to external components are depicted in Fig. \ref{fig_sysarch}.
The UDP interface to the host is realized by a custom-designed module, which supports full Gbit-Ethernet line speed.
All other modules on the FPGA firmware can be configured from the host via connections to individual UDP ports.
Each NeuroSoC chip has its corresponding chip control module in the FPGA that forwards configuration via the joint test action group (JTAG) protocol and sends/receives pulse packets.
Pulse routing, stimulation and tracing is provided by a central pulse routing module, which stores its routing information in an external double-data rate (DDR) random-access memory (RAM) included on the FPGA evaluation board.
An experiment control module allows for synchronous start and stop of experiments, and provides a global time base with a fixed resolution of 0.1ms.

Routing of pulses during an experiment works as follows:
In the NeuroSoC interface module, incoming pulses from a NeuroSoC are registered with the current global time.
Each pulse is subsequently duplicated four times.
For each of the duplicated pulses, an individual, configurable delay value can be added to the pulse time.
Thus, the system supports four independently configurable axonal delays per neuron.
Having calculated the target time, pulses are stored in a buffer inside each NeuroSoC interface module.
Once their target time is reached, pulses are sent to the routing module.
There, the information on the target neurons for each pulse is fetched from the external DDR RAM.
From there, pulses are sent to their targets immediately.
Each pulse can be routed to a maximum of 3.5k targets, which is enough for constructing arbitrary network topologies up to fully-connected networks.
The throughput of the whole routing chain is mainly limited by the input bandwidth per chip of 25~Mevent/s, corresponding to a 225~Mevent/s peak rate of synaptic events for the whole system.

The whole setup is controlled from a host PC via a combined C++/Python software stack, implementing a back-end for the PyNN 0.8 common simulator interface \cite{davison09}.
This allows for interoperability of the code with software simulators.
The back-end supports a standard conductance-based leaky integrate-and-fire neuron, as well as an extra neuron type for giving access to all five available synapse types.
A separate neuron type is employed for representing Ethernet connections to external setups \cite{rast13,george15}, allowing for seamless integration of remote setups in the PyNN script.
In particular, this enables real-time interaction with biological setups \cite{levi12}, making hybrid integration of neuromorphic hardware and biological neurons possible in-situ as well as remotely.

\subsection*{Mean-Field Approach}

\begin{figure}
\centering
\includegraphics[width=0.5\textwidth]{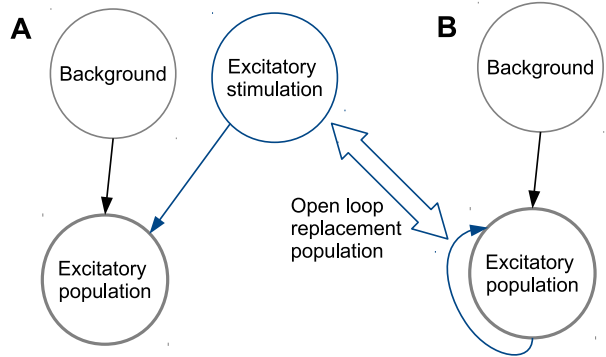}
\caption{\label{fig_network_structure} (A) Network structure for open-loop characterization; (B) Network structure for bursting modes.}
\end{figure}

We try to achieve bursting behavior with the simplest possible network model, taking a single population of excitatory neurons with recurrent coupling, and a background population that provides Poisson input.
As shown in Results, this is a sufficient configuration for a wide variety of bursting regimes.

For characterization, the recurrent connections are cut and replaced by a second Poisson input population, as shown in Fig. \ref{fig_network_structure}A.
This 'open loop' configuration allows to measure the transfer curve of the neuron population, i.e. the reaction to its own stimulation by the recurrent connections.
At all intersections of the transfer curve with the unity gain curve, the self-consistency condition of population output being equal to input from the recurrent coupling is fulfilled, denoting possible fixed points of the system.
Once the recurrent connections are closed (see Fig. \ref{fig_network_structure}B), the network will move towards one of its stable fixed points.
Transitions between fixed points may be triggered by fluctuations in the background population, by external stimulation, or by a dynamic change of the transfer curve.

\begin{figure}
\centering
\includegraphics[width=0.5\textwidth]{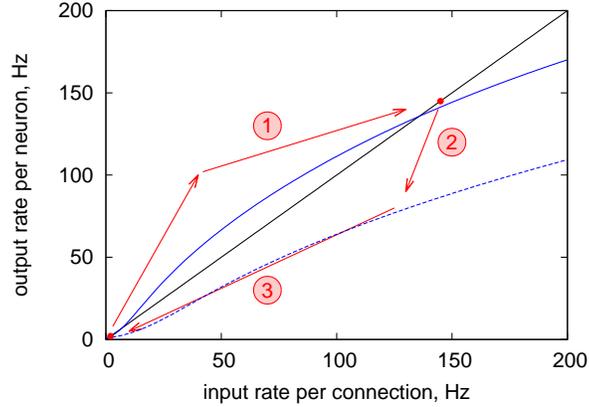}
\caption{\label{fig_transfer_theory} Sketch of transfer curves without (solid line) and with (dashed line) spike-frequency adaptation. The numbers and arrows show the process of burst generation, see text.  }
\end{figure}

The relation of the transfer curve with the network's operating points can be utilized for guiding the parameter tuning of the network.
For the pursued bursting behavior, several conditions need to be fulfilled in the transfer curve.
Bursting is a bi-stable operation, calling for two initially stable fixed points.
This is achieved by an S-shaped transfer curve, as shown in Fig. \ref{fig_transfer_theory} (solid line).
The upper and lower fixed points are stable, whereas the fixed point in the middle is instable, forming the boundary between the attracting regions of the two stable fixed points.

Initially, the network is in the low-rate stable fixed point.
Temporal variations in the Poisson background result in perturbations of the network around this point.
A network burst is initiated by a transition to the high-rate stable fixed point (see number 1 in Fig. \ref{fig_transfer_theory}).
This transition may happen spontaneously if the perturbations due to the Poisson background are big enough to bring the network beyond the boundary of the attracting region, i.e. the instable fixed point.
For bursting behavior, the network must move back to the low-rate stable fixed point spontaneously after a short time.
For this to happen, the high-rate fixed point must become instable, corresponding to the disappearance of the upper intersection with the unity gain curve (numbers 2 and 3 in Fig. \ref{fig_transfer_theory}).
This can be achieved by some form of inhibiting adaptation, damping the gain of the transfer curve (see dashed line).
For this purpose, we use spike-frequency adaptation. 
The length of a burst depends on how fast the spike-frequency adaptation builds up, which can be scaled by its amplitude.
In turn, the minimum inter-burst interval is related to the adaptation time constant, because only when the adaptation has decayed sufficiently, the bi-stable transfer curve is restored.
In absence of external stimulation, the time of the next burst depends on the interplay between the variance of the background noise and the shape of the transfer curve at low frequencies, as discussed above.
This can be thought of in terms of a transition rate from low- to high-rate fixed point.
If the transition rate is high, the next burst will happen shortly after sufficient decay of the adaptation, resulting in a regular bursting regime with a burst interval close to the minimum.
If the transition rate is lower, burst initiation becomes less probable per unit time, resulting in a more irregular bursting regime with higher mean inter-burst interval.
The shape of the transfer curve at low rates, and thus the transition rate, can be influenced effectively by the amplitude of recurrent connections, changing the overall gain of the transfer curve.
Therefore, we expect a strong dependence of the bursting regime on this amplitude.

For avoiding synchronization in the activities of single neurons, we chose a sufficiently big network with sparse connectivity.
Specifically, we employed all 2880 neurons available in system for the excitatory population and used random connectivity, where the connection probability was set such that each neuron receives on average 20 recurrent and 20 background connections.
This network configuration also makes a mean-field calculation of the transfer curves applicable, which we use for comparison with the measured transfer curves in the Results section.
Details on the employed mean-field approximation can be found in the Appendix.

Following the above approach, we first tuned the transfer curves without adaptation, resulting in the parameters listed in Table \ref{tab_parameters}.
Afterwards, we added spike-frequency adaptation, choosing the maximally possible time constant of 330ms for restricting the maximum burst frequency.
The two remaining free parameters are the conductance of recurrent connections and the conductance amplitude for adaptation.
Measured transfer curves with the chosen parameters are detailed in the Results section.

\begin{table}
\centering
\caption{\label{tab_parameters} List of network parameters}
\footnotesize
\begin{tabular}{|ll|p{2.5cm}|}
\hline 
resting potential & $v_\mathrm{rest}$ & -65mV \\
reset potential & $v_\mathrm{reset}$ & -80mV \\
threshold potential & $v_\mathrm{thresh}$ & -50mV \\
membrane capacitance & $C_\mathrm{mem}$ & 1nF \\
membrane time constant & $\tau_\mathrm{mem}$ & 8ms \\
refractory period & $T_\mathrm{refrac}$ & 2.5ms \\
synaptic time constant & $\tau_\mathrm{syn}$ & 8ms \\
synaptic reversal potential & $E_\mathrm{syn}$ & 0mV \\
adaptation time constant & $\tau_\mathrm{sfa}$ & 330ms \\
adaptation reversal potential & $E_\mathrm{sfa}$ & -80mV \\ \hline

neurons in network & $N$ & 2880 \\
external background sources & $N_\mathrm{bg}$ & 200 \\
probability of recurrent connections & $p_\mathrm{rec}$ & 0.007(=20/$N$) \\
probability of background connections & $p_\mathrm{bg}$ & 0.1(=20/$N_\mathrm{bg}$) \\
conductance of background connections & $\hat g_\mathrm{bg}$ & 5nS \\
rate per background source & $f_\mathrm{bg}$ & 16Hz \\
\hline
\end{tabular}
\end{table}

\subsection*{Analysis Methods}

For characterizing the bursting behaviour of a network, the following procedures and measures were used.
For each parameter set, an experiment of 500s was run.
The resulting output spikes were divided in bins of 50ms.
From the number of spikes per bin, the mean firing rate per neuron was calculated for each bin.
A burst was detected if the mean firing rate per neuron exhibited a value of more than 20Hz in one or more bins.
All subsequent bins above that threshold were counted as one burst, i.e. the next bin below 20Hz would be detected as the end of the burst.

Two measures were used for network characterization, the burst length and the inter-burst interval (IBI).
The burst length was taken as the number of subsequent bins above 20Hz.
For characterization, mean and coefficient of variation (CV) were calculated over all detected bursts.
Likewise, the IBI was taken as the number of subsequent bins below 20Hz.
Again, mean and CV over all detected IBIs were used for characterization.
These statistical measures were only calculated for those runs, were the number of detected bursts was greater than 50, for the results to be informative.

\section*{Results}

\subsection*{Open-Loop Network Measurements}

Following the mean-field approach described in Materials and Methods, we first characterize the bursting network via open-loop measurements, which allows to predict its dynamical behavior in the final recurrent setting.
Building upon mean-field theory not only guides the parameter tuning process, but also allows for a direct comparison of theory and measurement results.

Before assessing the final network model, we first characterize the hardware variations with deterministic stimulation and connectivity.
For this, we modify the network definition given in Tab. \ref{tab_parameters}:
The number of synaptic inputs to each neuron is fixed to each 20 for background and recurrent projections, and the Poisson stimulation is generated with a fixed seed.
As a result, each neuron is parameterized identically and stimulated independently with the same spike train.

\begin{figure}
\centering
\includegraphics[width=0.45\textwidth]{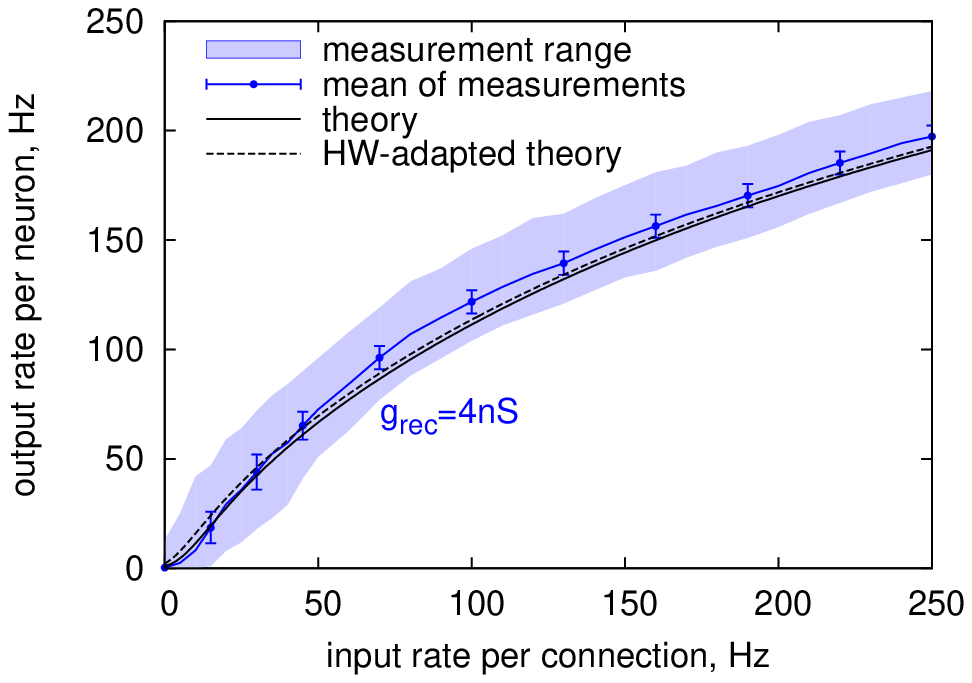}
\includegraphics[width=0.45\textwidth]{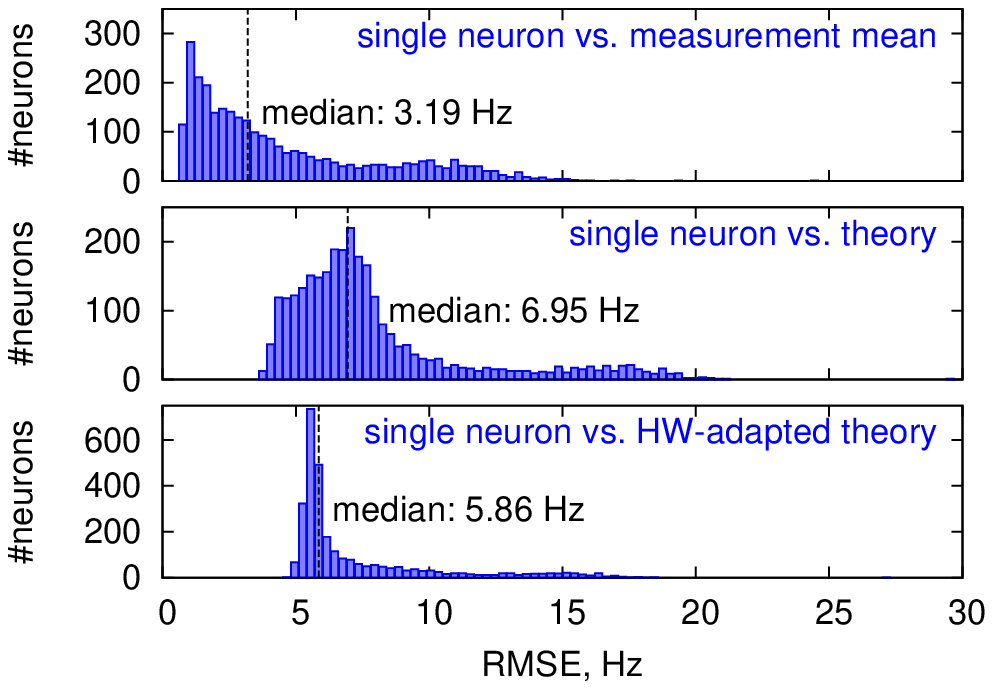}
\caption{\label{fig_transfer_single} Measured open-loop transfer curves of single neurons. Left: Mean transfer curve and deviations compared to mean-field theory. The error bars denote the 1$\sigma$-interval over all measurements. Right: Histograms for deviations of single measurements from measurement mean and theoretically predicted transfer curves.}
\end{figure}

Figure \ref{fig_transfer_single} shows summarized results of the single-neuron measurements.
Variations in the transfer curves are relatively low, with a median RMSE from the mean of 3.19 Hz.
Only a few outliers exhibit significantly higher RMSE, expanding the range of measurements.
Comparing the measurements with the theoretical mean-field approximation (cf. solid line) shows a slight systematic deviation at higher frequencies.
This also makes deviations of single measurements from the theoretical prediction higher (median RMSE of 6.95 Hz).

A reason of this systematic deviation may be in the switched-capacitor circuit principle, resulting in discrete switching events on the neuron membrane.
In turn, the statistical variation of the membrane voltage may be affected, which would have an impact on the average spiking behavior of the neurons.
To estimate the impact of this effect, we adapted the mean-field approximation to account for this effect, resulting in a modified formulation for the variance of the membrane voltage (see Appendix for details).
The hardware-adapted transfer curve only marginally deviates from the original mean-field prediction (cf. dashed line), but slightly improves matching with the hardware measurements at high frequencies, resulting in a median RMSE of 5.86 Hz.

Despite these variations, the measured transfer curves are in good agreement with the mean-field approximation.
It has to be emphasized that these results were achieved without any calibration or tuning of the transformation from model to hardware parameters.
Reduced deviations could be achieved by tuning single parameters of the hardware, e.g. the synaptic strength, but this would deviate from our idea of an easily and generally applicable hardware system.
With these results, we move on to the open-loop characterization of the actual network model.

\begin{figure}
\centering
\includegraphics[width=0.45\textwidth]{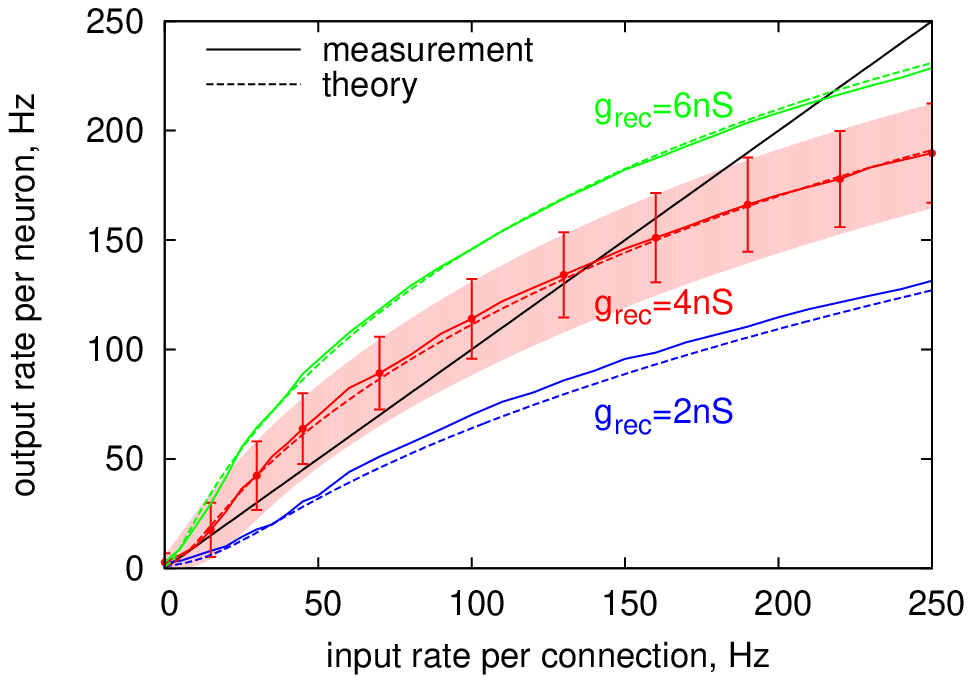}
\includegraphics[width=0.45\textwidth]{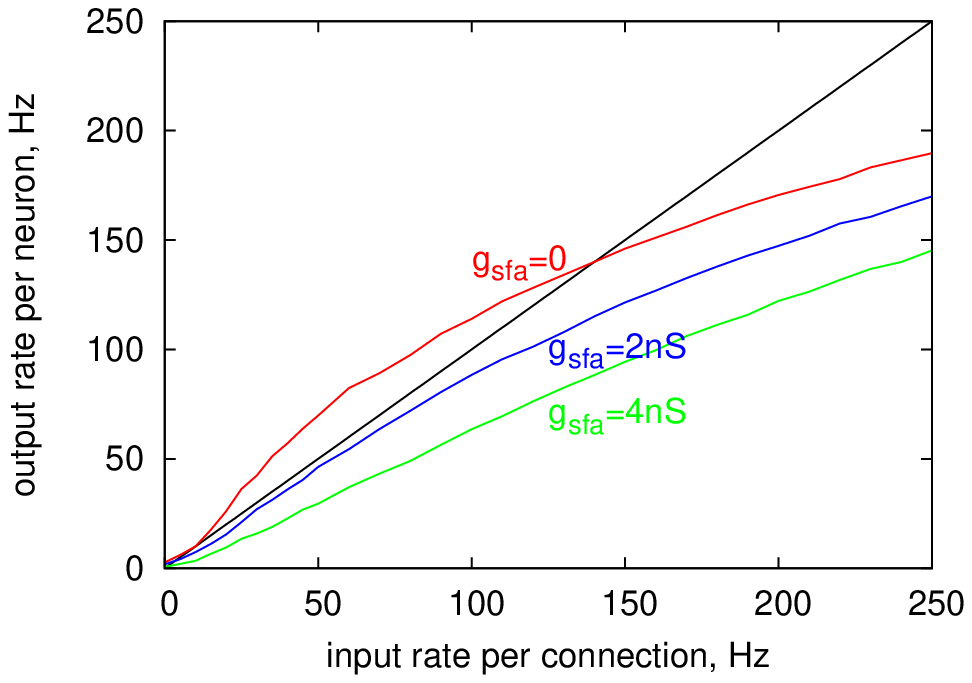}
\caption{\label{fig_transfer} Measured open-loop transfer curves with varying $g_\mathrm{rec}$ (left) and $g_\mathrm{sfa}$ (right). For each measured input frequency, the network was stimulated for 2 seconds, and the firing rate was computed for each neuron separately, discarding the first second to eliminate the influence of transient effects. An 0.5 second phase without stimulation separated the single measurements, letting the network relax to a resting state. Solid lines show the measured mean over all neurons, while dashed lines denote the mean-field approximation for the same parameters. The error bars depict the variation in the single neuron responses (1$\sigma$-interval) for the case $g_\mathrm{rec}=4$nS,  $g_\mathrm{sfa}=0$. The shaded area shows the range of mean-field approximations with varying number of synapses per neuron due to the employed random connectivity (1$\sigma$-interval of synapse count) for the same parameters.}
\end{figure}

Figure \ref{fig_transfer} shows measured open-loop transfer curves for the network with all parameters as in Tab. \ref{tab_parameters}.
The transfer curves exhibit an S-shape, as required for a bistable closed-loop behavior.
A stable high-rate state develops at a conductance between $g_\mathrm{rec}=2$nS and $g_\mathrm{rec}=4$nS for recurrent connections.
Measurements and theoretical predictions for the mean show only minor deviations.
Compared to the single-neuron measurements, deviations are even smaller, with an RMSE of measured $g_\mathrm{rec}=4$nS curve to the mean-field prediction of 2.09 Hz.
One reason for this is the averaging over different synapse counts in the employed network, which slightly dampens the response at higher rates compared to a fixed synapse count, partially compensating for the deviations present in the single-neuron measurements.

The highest deviation occurs for the $g_\mathrm{rec}=2$nS curve at high rates.
A simple constant-current approximation generally showed better correspondence with the measurement results at high rates, also for this case.
However, it failed to explain the behavior at low input rates.
Still, the specific deviation for $g_\mathrm{rec}=2$nS might be an effect of the employed mean-field approximation.

Compared to these differences, the variance in the single-neuron transfer curves is relatively high in the employed network, see error bars and shaded area.
In particular, it is much higher than the variations seen in the previous single-neuron characterization (cf. Fig. \ref{fig_transfer_single}).
This effect can be well explained by the spread in the number of synapses per neuron due to random connectivity, and the corresponding differences in total synaptic activation.
Again, the variances in the measured curves well match with those of the theoretical predictions.
This observation confirms the conclusion from the single-neuron characterization that circuit mismatch has no significant impact on the network behavior on the population level.

The influence of the neuron adaptation is as expected (see right plot in Fig. \ref{fig_transfer}).
It lowers the transfer curve, the S-shape with two stable fixed points disappears.
At a sufficient strength $g_\mathrm{sfa}$ of the adaptation, the high-rate state gets instable.
When progressing from a non-adapted to an adapted state, this would force the network back to a low-rate state.
With higher amplitude $g_\mathrm{sfa}$, the distance to the unity gain curve gets higher, which predicts that the high-rate state collapses faster.

With these measurement results, all prerequisites in the transfer curves for bursting behavior, as sketched in Materials and Methods, are fulfilled.
We can therefore now go on to measuring the closed-loop behavior.

\subsection*{Closed-Loop Network Measurements}

\begin{figure}
\centering
\includegraphics[width=0.9\textwidth]{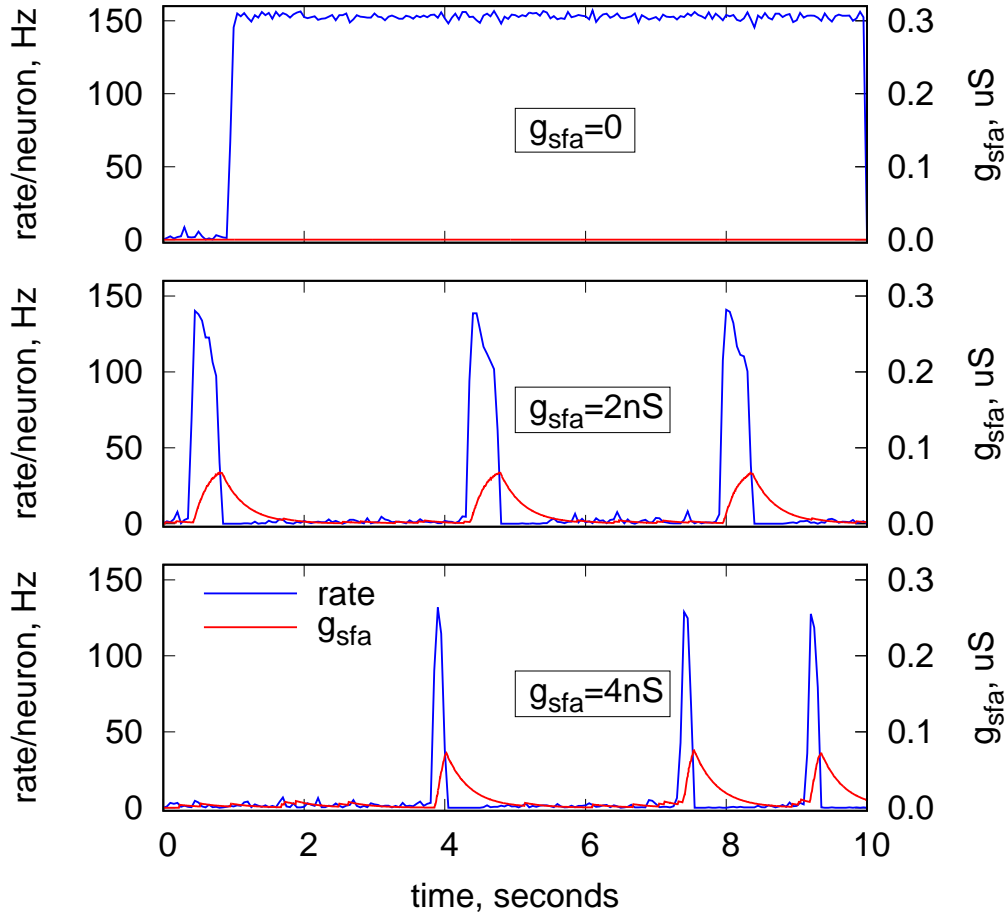}
\caption{Change of network behaviour with increasing influence of SFA.}
\label{fig_transient_sfa}
\end{figure}

In this section, we analyze the behaviour of the closed-loop network setting.
Figure \ref{fig_transient_sfa} shows the frequency behaviour of the excitatory population for increasing levels of self-adaptation.
All the other parameters, including the efficacy of the recurrent connections and the external stimulation are held constant throughout the trials.
At the beginning of each trial the network is prepared in a state of low activity and then it evolves autonomously for 10 seconds.

The trial reported in the upper plot shows the network output in the absence of adaptation.
The network exhibits a bistable behaviour:
It starts from a low firing rate at about 2Hz up to 0.9s when it abruptly jumps to the upper state, where the average firing rate is about 150Hz.
The average firing rate of the two states are in good agreement with the predictions of the open-loop transfer function described in the previous section.
Both the lower and the upper state are meta-stable states of the dynamics and the jump between the two is due to an instantaneous fluctuation of the neuronal firing rate.
Such fluctuations are due to the Poissonian nature of the external input and to the so called finite-size noise endogenous of a spiking network of sparsely and randomly connected neurons \cite{amit97,mattia04}.
The former source of noise is dominant when the network is in the lower state, while the latter dominates when the network is in the upper state.
Balancing the noise levels, to gain control over the bistable network, is a tricky point, especially in a neuromorphic mismatched network endowed with a massive amount of positive feedback, as described in \cite{giulioni12}.
Here, we follow the same approach of employing the open-loop transfer function, while the process is simplified due to the limited mismatch of the underlying switched-capacitor circuits and the good correspondence of theory and measurements.

Starting from a controlled noisy bistable behaviour, we progressively add self-inhibition (SFA), as shown in the middle and lower plot of Fig. \ref{fig_transient_sfa}.
It is evident that the increase in SFA causes a reduction of the average up-state duration.
In the lower plot, the self-inhibition level is such that the up-states are completely unstable.
In this condition, as soon as the network tries to jump up, it is immediately kicked back by the inhibition.
The result is a bursting behaviour of the network.
In this conditions bursts do not last more than 200ms reaching a maximum frequency of 140Hz.
We stress here that this behaviour is possible only thanks to the presence of noise, which makes the lower level a meta-stable one.
By changing a single parameter, we are able to move from a bistable to a bursting network behaviour.

To better understand this transition we should consider the two coupled dynamics: the neuronal one and the SFA one.
Mathematical and numerical analysis of this double-dynamics have been reported in \cite{mattia12,gigante15}.
Intuitively, observing the dynamics at the population level, we can state that the fast neuronal dynamics drives the slower SFA which in turn, with a certain delay, inhibits the neuronal activity.
This mechanism is evident in the middle plot of Fig. \ref{fig_transient_sfa}, and we can divide it into 3 phases.
In the first phase (from 2s to 4s) the network is in the lower state, the SFA level is practically zero and it does not affect the neuronal dynamics.
The second phase starts with a random fluctuation which induces a transition towards the upper state (4.3s).
The SFA conductance slowly increases and the inhibitory effect starts destabilizing the upper state.
The third phase starts with a downward transition (4.7s).
From this point on, the SFA conductance decays slowly ensuring a time period in which a new upward transition is unlikely to happen.
When the SFA level is sufficiently low we are back in the first phase and this cycle can start again triggered by a new noisy fluctuation. 

The duration of the various phases clearly depends on the time scales of the neuronal and SFA dynamics, their relative strength, and on the level of noise in the network.
In the example in Fig. \ref{fig_transient_sfa} we vary the increase of the SFA conductance by an individual spike, $g_\mathrm{SFA}$, from 0 to 4ns.
Qualitatively, varying $g_\mathrm{SFA}$ we have two different scenarios.
For low levels of $g_\mathrm{SFA}$, even when the SFA conductance reaches its maximum, two meta-stable states are still allowed.
In this case the self-inhibition simply slightly destabilizes the up-states such that their average duration is reduced.
For high values of $g_\mathrm{SFA}$, the upper meta-stable state of the dynamics exists only if the SFA conductance is sufficiently low.
In this condition, when the SFA conductance is almost zero the network can jump towards the upper stable state, which however ''disappears'' during the transition, due to a fast increase of the SFA level (cf. also the corresponding transfer curves in the right plot of Fig. \ref{fig_transfer}).
Dependent on the exact moment at which the SFA conductance reaches the critical threshold, the transition can not be initiated at all or not completed.
The latter condition is the one shown in the lower plot of Fig. \ref{fig_transient_sfa} where the network creates short bursts of high activity.

\begin{figure}[ht]
\centering
\includegraphics[width=0.4\textwidth]{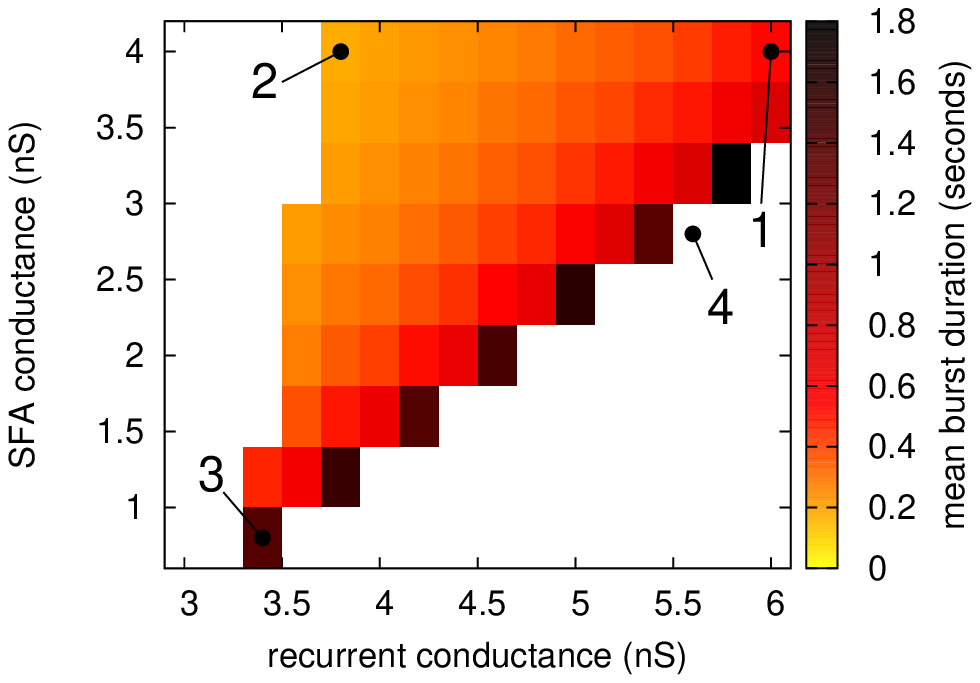}
\includegraphics[width=0.4\textwidth]{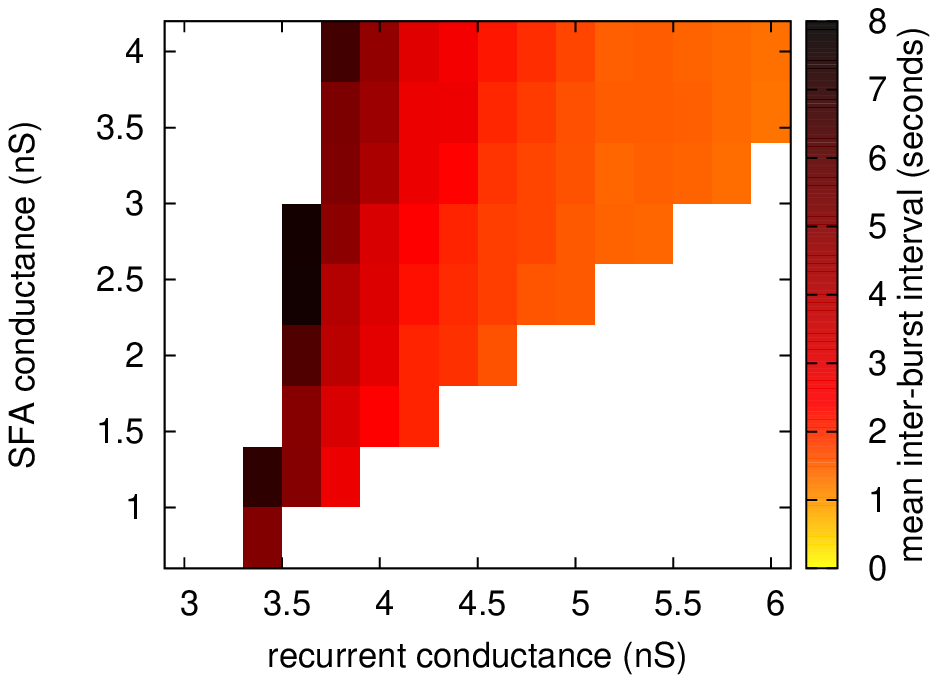}

\includegraphics[width=0.4\textwidth]{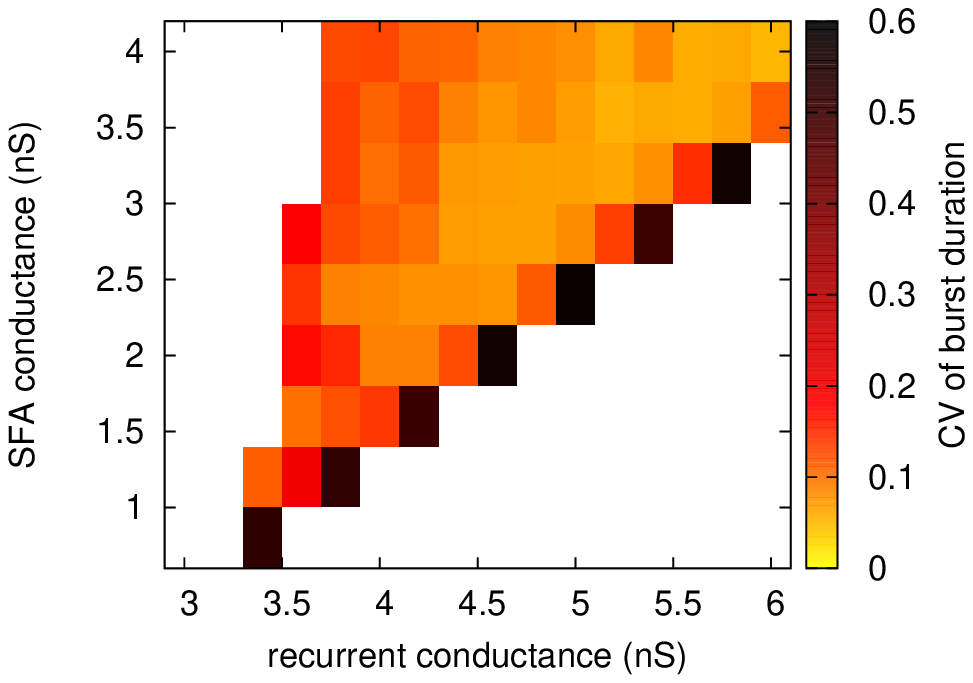}
\includegraphics[width=0.4\textwidth]{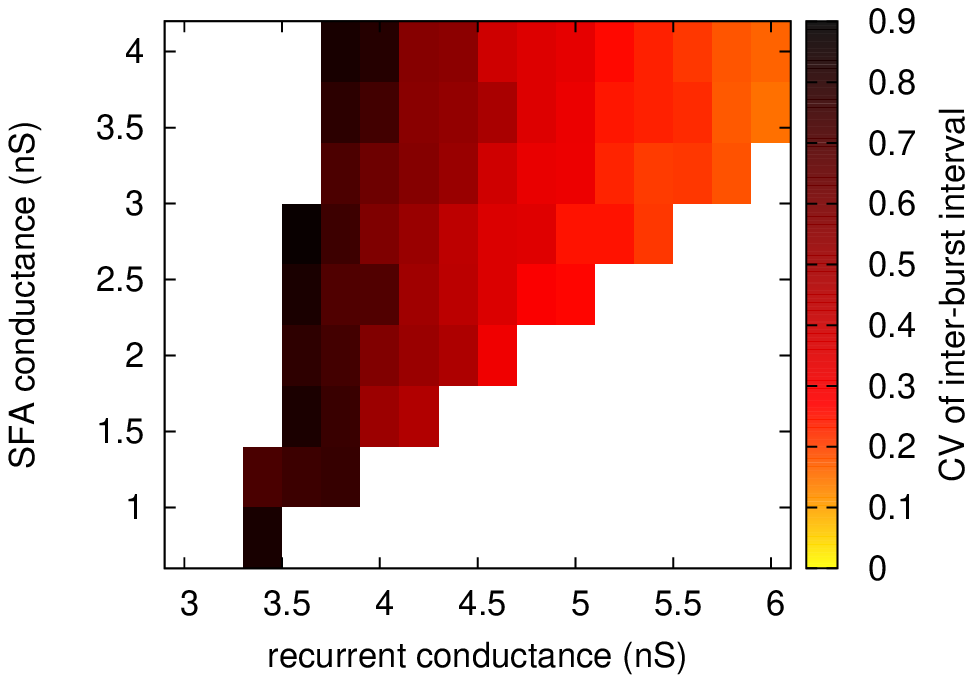}

\includegraphics[width=0.6\textwidth]{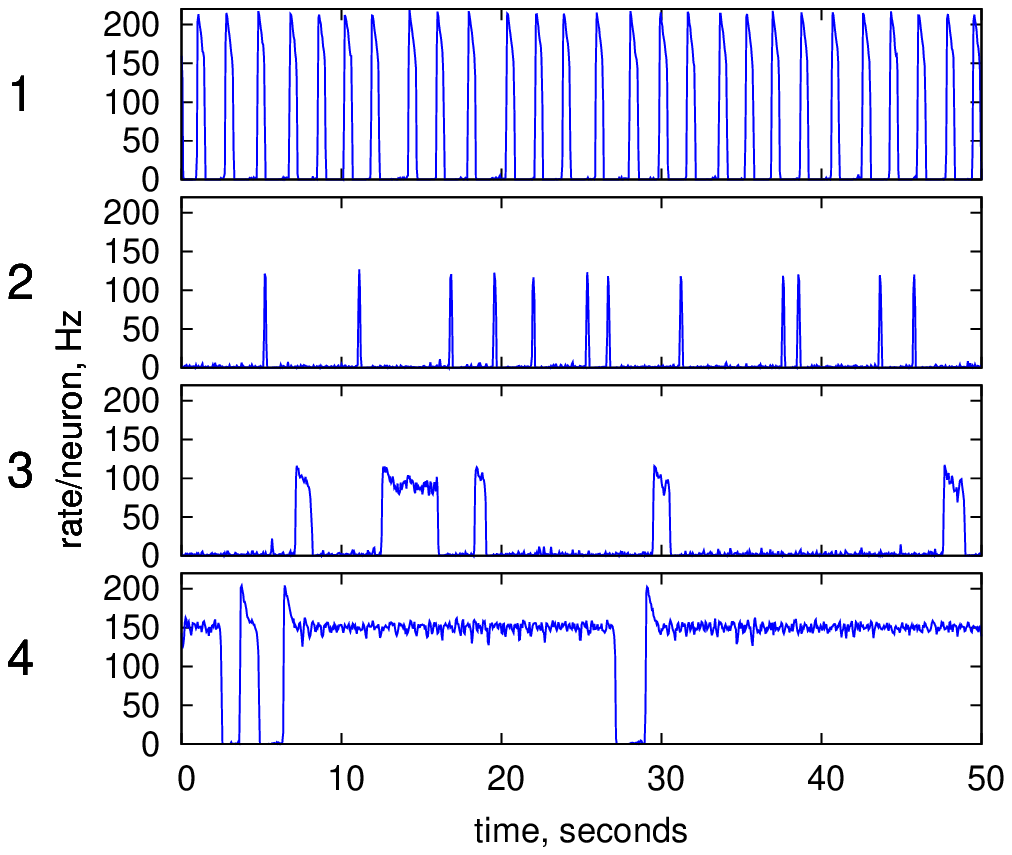}

\caption{Phase-plane of bursting behaviour, varying the efficacies of recurrent connections, $g_\mathrm{rec}$, and self-inhibition, $g_\mathrm{SFA}$. The transients in the lower half of the figure illustrate network behaviour at different positions in the phase-plane, as denoted in the upper left plot. }
\label{fig_phaseplane}
\end{figure}

The SFA increase per spike $g_\mathrm{SFA}$ is only one dimension of the parameter space, and the duration of the up-states only one of the interesting characteristics of the dynamics.
In this paragraph we explore the phase plane $g_\mathrm{SFA}$ vs recurrent coupling strength $g_\mathrm{rec}$.
For each couple $(g_\mathrm{SFA}, g_\mathrm{rec})$, we ran one experiment with a duration of 500s.
We analyzed the distribution of the burst durations and the distribution of the inter-burst-intervals (IBI).
The mean and the coefficient of variation of those distributions are reported in Fig. \ref{fig_phaseplane}.
The different network behaviours, over this phase-plane, result from different equilibriums between two ''forces'': the tendency of jumping to the upper state, which increases with $g_\mathrm{rec}$, and the tendency of destabilizing the upper state, which increases with $g_\mathrm{SFA}$.

In the upper right corner of the plane both those forces are strong, and the result is the quasi deterministic oscillation shown in the transient plot number 1.
In this point of the phase-plane the time-scale of the dynamics is mostly governed by the SFA.
The duration of the upper state is governed by the ratio of $g_\mathrm{SFA}$ to $g_\mathrm{rec}$, and upward transitions happen soon after the release of the self-inhibition, on average after 5.5$\tau_\mathrm{SFA}$.

Moving leftwards, $g_\mathrm{rec}$ reduces, and the tendency of jumping up decreases.
Hence the lower point of the dynamics becomes more and more stable and the average IBI increases.
Also the CV of the IBI increases since larger fluctuations are now necessary to trigger upward transitions.
In other words, moving leftwards we are moving towards a more noise-driven regime (see transient plot number 2).
We note here that the IBI distributions are poorly affected by $g_\mathrm{SFA}$.
This is coherent with the fact that in the lower state the self-inhibition is negligible a few $\tau_\mathrm{SFA}$ after the last downwards transition.
On the contrary, the effect of the SFA is relevant in the upper state.
Therefore, the average burst duration decreases both at increasing $g_\mathrm{SFA}$ and decreasing $g_\mathrm{rec}$, since the stability of the upper state reduces in both cases.

Moving downwards on the phase-plane the force destabilizing the upper state decreases.
Hence we have longer up-states lasting up to a sufficiently high noise fluctuation, which also implies a higher CV of their duration.
When $g_\mathrm{SFA}$ is below a certain level relative to $g_\mathrm{rec}$, up-states become dominant and the network leaves its bursting regime, as shown by the transient plots 3 and 4.
This transition is quite sharp, and its position can be well described by a linear relation between $g_\mathrm{SFA}$ and $g_\mathrm{rec}$.

These results demonstrate the level of fine control that can be reached by a theory-driven approach to the tuning of network dynamics, together with a hardware implementation strategy resulting in limited mismatch effects.

\section*{Discussion}

\subsection*{Neuromorphic system:} 

From its inception in 1989 \cite{mead90}, neuromorphic engineering tried to mimic the design and operating principles of neural networks, to develop biomimetic microelectronic devices which implement biological models \cite{bartolozzi07}. So far, the neuromorphic approach has been successful in implementations of sensory functions (e.g. visual processing \cite{koenig02}) and computational functions that rely on building blocks of brain processing (e.g. pattern recognition \cite{qiao15}). 
Here, we present a neuromorphic system intended for use in a biohybrid, i.e. coupled to a cultured in-vitro network. The neuromorphic system is optimized for biologically realistic short-term dynamics, carried out in switched capacitor (SC) technique. We have previously shown that using SC a system can be implemented in ultra-deep submicron CMOS \cite{mayr15}, with synapse density on par with modern nano-scale approaches \cite{du15,mostafa15}. Here, we show that SC also makes for a very reproducible system behaviour, which enables the construction of large-scale neuromorphic systems as reproducible behaviour significantly eases configurability. 

The intended usage as neuromorphic biohybrid only necessitates the replication of short-term dynamics. Thus, the system completely omits long-term plasticity \cite{fusi00,mayr10b}, enabling the use of multisynapses and a corresponding increase in network size, as per-synapse state variables are not required. Table \ref{tab_characteristics} gives an overview of the chip and overall system characteristics. System size is among the larger systems deployed today, with a competitive power consumption. The system size was dictated by the requirement to enable a network size of several 1000 neurons with dense connectivity to act as credible counterpart to a petri dish culture with a similar number of neurons. The system implements a variety of different biophysical mechanisms, such as conductance-based GABA, AMPA, NMDA synapses after the models in \cite{rolls10} and different types of presynaptic adaptation derived from \cite{markram98}. Great care was taken to faithfully emulate these models in their biological richness \cite{noack14}. 

\begin{table}
\centering
\caption{\label{tab_characteristics}Characteristics of the presented SC neuromorphic chip (in brackets: overall system).}
\footnotesize
\begin{tabular}{|p{5cm}|p{5cm}|}
\hline 
Technology & UMC 180~nm \\
\hline
Number of neurons & 320 (2880)\\
\hline
Number of hardware synapses & 1.6k (14.4k)\\
\hline
Number of virtual synapses & 16M (144M) \\
\hline
Number of presynaptic adaptation circuits & 1280 (11520) \\
\hline
Chip area & 5*10~mm \\
\hline
Supply voltage & xx \\
\hline
Power consumption & xx \\
\hline
Energy/spike & xx \\
\hline
\end{tabular}
\end{table}

\subsection*{Mesoscopic characterization}

We used the well-established mean-field approach 
for parameter tuning of the employed network model. For this, we characterized the recurrent network by its open-loop transfer function, ensuring that it exhibited the required features for bursting behavior, such as an S-shape and a sufficiently strong inhibitory adaptation.
The low impact of device mismatch on the switched-capacitor circuits resulted in a very good agreement of the measurements with the expected behavior from mean-field theory without calibration or problem-specific tuning.
This is an advantage compared to existing mixed-signal neuromorphic systems, where elaborate hardware calibration was performed before operation, 
or the hardware transfer curve was directly tuned without quantitative correspondence to a given mean-field approximation \cite{giulioni12}.
As a consequence, no individual parameter storage per neuron is required for calibration, but parameters can be stored in groups, reducing silicon area for storage, in our case by a factor of 32.
Furthermore, reproducing mean-field theory not only qualitatively, but quantitatively significantly reduces the effort to utilize concrete theoretical results in neuromorphic hardware.
Moreover, with this level of correspondence, neuromorphic hardware can be employed as a direct real-time test bed for theoretical predictions.

Already with a relatively simple network model, we showed a wide range of bursting behavior in hardware. 
With our results, the dynamical behavior of the network, characterized by burst length, inter-burst interval and their distribution, can be tuned by choosing a suitable combination of only two parameters.
These two, the strength of recurrent connections and the strength of adaptation, are key for achieving a desired behavior.
However, further parameters can be utilized for more extended control.
For example, the strength of the background input influences the probability of burst initiation and the adaptation time constant defines the minimum inter-burst interval.
For more complex behavior, the other hardware features, such as synaptic short-term adaptation and NMDA synapses can be employed, and a more elaborate network model could be chosen.
However, such extensions should always be guided by a theoretical framework, avoiding 'blind' parameter tuning without an understanding of the underlying dynamical mechanisms. Our described mesoscopic tuning framework can easily be extended to using more complex adaptation mechanisms provided by the chip, as the basic mechanisms of tuning curves and bursting stay similar.


\subsection*{Hybrid Usage:} 

Bursting is a widespread mesoscopic phenomenon in biological networks.
Neuromorphic hardware behaving similarly is a prerequisite for a seamless dynamical integration with biological networks in hybrid systems. 
Here, we show only a subset of the neuromorphic functionality, i.e. excitatory AMPA synapses and Calcium-modulated postsynaptic adaptation. With this limited functionality, we already achieve complex, tunable dynamics. Analysis of the future biohybrid interface to nerve cells will show which mechanisms we need to further enable on the chip to achieve a seamless coupling of dynamics between petri dish culture and neuromorphic network.
In extension of the presented work, our network would need an additional bursting input, such as indicated in Fig. \ref{fig_network_structure}, and a complementary analysis of its dynamical behavior would be required.
With this additional bursting input provided by biology, the neuromorphic hardware network could work as an extension or partial replacement of biological tissue, forming a recurrent hybrid network.
The detailed controllability of the hardware network has the potential for a more fine-grained and natural interaction with biological networks.
Also, it offers a powerful tool for better understanding the behavior of modular networks, 
utilizing the simpler and finer adjustability of neuromorphic hardware compared to biological networks.


\section*{Acknowledgments}
This research has received funding from the European Union Seventh Framework Programme (FP7/2007- 2013) under grant agreement no. 269459 (CORONET)


%
%
%

\section*{Appendix}

For the employed leaky-integrate-and-fire neuron with spike-frequency adaptation (SFA), the membrane voltage $v$ progresses dependent on synaptic input current $i_\mathrm{syn}$ and adaptation current $i_\mathrm{sfa}$ as:
\begin{equation}
  C_\mathrm{mem}\frac{dv}{dt} = g_\mathrm{mem}\cdot (v_\mathrm{rest}-v) + i_\mathrm{sfa} + i_\mathrm{syn}\;,
\end{equation}
where $v$ is the membrane voltage, $C_\mathrm{mem}$ and $g_\mathrm{mem}$ are membrane capacitance and conductance, and $v_\mathrm{rest}$ denotes the resting potential.
The resulting membrane time constant is given by $\tau_\mathrm{mem}=C_\mathrm{mem}/g_\mathrm{mem}$.
Each time $v$ reaches the threshold voltage $v_\mathrm{thresh}$, it emits a spike and the membrane voltage is held at $v_\mathrm{reset}$ for the refractory period $T_\mathrm{refrac}$.
As detailed in the chip description, synaptic input is generated by conductance-based synapses with exponentially decaying conductances $g_{\mathrm{syn},i}$, which can be formulated as:
\begin{align}
 i_\mathrm{syn} &= \sum_{i} g_{\mathrm{syn},i}\cdot (E_{\mathrm{syn},i}-v) \\
 \tau_{\mathrm{syn},i}\frac{dg_{\mathrm{syn},i}}{dt} &= -g_{\mathrm{syn},i} + \hat g_{\mathrm{syn},i} \sum_{k}\delta (t-t_{i,k})
\end{align}
In this equation, $E_{\mathrm{syn},i}$ is the synaptic reversal potential, $t_{i,k}$ is the time of the $k$-th spike at synapse $i$, and $\hat g_{\mathrm{syn},i}$ denotes the strength of synapse $i$.
The spike-frequency adaptation current $i_\mathrm{sfa}$ is described as an inhibitory conductance-based synapse ($\hat g_{\mathrm{syn},i}<0$), driven by the spikes of the postsynaptic neuron.

We want to calculate analytical transfer curves for parameter tuning and comparison to measurement results.
For this, we adapted the mean-field calculations detailed in \cite{renart03}.
We did not incorporate spike-frequency adaptation, as it is utilized in our network only as a transient effect to finish a network burst.
In contrast, the following mean-field calculation derives a steady-state solution.

Following the mean-field approach, all neurons are assumed to be statistically equivalent.
In the employed network, all excitatory synapses have time constant $\tau_\mathrm{syn}$ and reversal potential $E_\mathrm{syn}$.
Strength of synapses, $\hat g_{\mathrm{syn},i}$, is $g_\mathrm{rec}$ for recurrent connections and $g_\mathrm{bg}$ for connections from background.
According to \cite{renart03}, the mean firing rate per neuron $f_\mathrm{out}$ can be approximated by:
\begin{equation}
 \frac{1}{f_\mathrm{out}} = T_\mathrm{refrac} + \tilde\tau_\mathrm{mem}\sqrt{\pi} \int_{\frac{v_\mathrm{rest}-v_\mathrm{ss}}{\sigma_v}}^{\frac{v_\mathrm{thresh}-v_\mathrm{ss}}{\sigma_v}}
    \mathrm{e}^{x^2} (1+\mathrm{erf}(x))dx\,,\;\mbox{with}\; \mathrm{erf}(x)=\frac{2}{\sqrt{\pi}}\int_0^x \mathrm{e}^{-u^2}du\;,
 \label{eq_meanfield}
\end{equation}
where $\tilde\tau_\mathrm{mem}$ is the effective membrane time constant considering synaptic conductances, $v_\mathrm{ss}$ is the membrane voltage in steady-state, and $\sigma_v$ is the standard deviation of the membrane voltage.

The effective membrane time constant $\tilde\tau_\mathrm{mem}$ results from the parallel connection of the membrane conductance $g_\mathrm{mem}=C_\mathrm{mem}/\tau_\mathrm{mem}$ and the total synaptic conductance $g_{\mathrm{syn,total}}$ \cite{renart03}:
\begin{equation}
 \tilde\tau_\mathrm{mem} = \frac{C_\mathrm{mem}}{g_\mathrm{mem} + g_\mathrm{syn,total} }\;.
\end{equation}
Here, $g_\mathrm{syn,total}$ is taken as the mean synaptic conductance for the current input rate $f_\mathrm{in}$ and the background rate $f_\mathrm{bg}$:
\begin{equation}
 g_\mathrm{syn,total} = \tau_\mathrm{syn}\cdot (g_\mathrm{rec}f_\mathrm{in}N_\mathrm{conn,rec} + g_\mathrm{bg}f_\mathrm{bg}N_\mathrm{conn,bg})\;.
\end{equation}
In this equation, $N_\mathrm{conn,rec} = N\cdot p_\mathrm{rec}$ and $N_\mathrm{conn,bg}=N_\mathrm{bg}\cdot p_\mathrm{bg}$ denote the average numbers of recurrent and background connections per neuron, respectively.
Please note that $f_\mathrm{in}$ and $f_\mathrm{out}$ both represent the mean firing rate of a neuron in the excitatory population, the former presynaptically, the latter postsynaptically.
This separation is a direct consequence of the open-loop characterization; valid fixed points of the recurrent network are those where $f_\mathrm{out}=f_\mathrm{in}$.

When the membrane voltage is at its steady-state value $v_\mathrm{ss}$, average currents through the synaptic conductances and the membrane conductance equalize, resulting in
\begin{equation}
 v_\mathrm{ss} = \frac{v_\mathrm{rest}g_\mathrm{mem} + E_\mathrm{syn}g_{\mathrm{syn,total}}}{ g_\mathrm{mem} + g_{\mathrm{syn,total}} }\;.
\end{equation}

Finally, following \cite{renart03}, the standard deviation $\sigma_v$ of the membrane voltage distribution can be approximated as
\begin{equation}
 \sigma_v = \sqrt{\frac{\sigma_i^2 \cdot \tilde\tau_\mathrm{mem}}{C_\mathrm{mem}^2}}\;,\;\mbox{with}\;\;
 \sigma_i^2 = f_\mathrm{in}N_\mathrm{conn,rec}Q_\mathrm{rec}^2 + f_\mathrm{bg}N_\mathrm{conn,bg}Q_\mathrm{bg}^2\;,
\end{equation}
where $Q_\mathrm{rec}=g_\mathrm{rec}\tau_\mathrm{syn}(E_\mathrm{syn}-\bar v)$ and $Q_\mathrm{rec}=g_\mathrm{bg}\tau_\mathrm{syn}(E_\mathrm{syn}-\bar v)$ are approximations for the average charge transported by a single incoming spike over a recurrent or background connection, respectively. For the average membrane voltage $\bar v$, we use $\bar v=(v_\mathrm{thresh}-v_\mathrm{reset})/2$.

For each frequency $f_\mathrm{in}$, calculating the frequency-dependent variables $\tilde\tau_\mathrm{mem}$, $v_\mathrm{ss}$, $\sigma_v$ and inserting them in Eq. \ref{eq_meanfield} yields an estimate for the output frequency $f_\mathrm{out}$.
A sweep over $f_\mathrm{in}$ then gives the transfer curve $f_\mathrm{out}(f_\mathrm{in})$.
If not noted otherwise, we use this approximation from mean-field theory for comparison with the hardware measurements.

The actual membrane voltage in the hardware neurons behaves slightly different from the above assumptions due to the switched-capacitor circuit technique.
As shown in Fig. \ref{fig:digital}, conductance changes for the same synapse type are accumulated in a digital register GSYN\_REG.
The value of this register translates into a switching frequency of a switched-capacitor circuit connected to the membrane capacitance.
Each switching cycle of this circuit results in a jump of the membrane voltage due to the charge equalization between the membrane capacitance $C_m$ and the respective synaptic or leak conductance $C_\mathrm{syn}$ or $C_L$.
The size of this jump is $\alpha\cdot (E_{syn}-v)$ and $\alpha\cdot (E_L-v)$, where $\alpha=C_L/C_m=C_{syn}/C_m=1/20$ is the constant capacitance ratio.
As a consequence, while the membrane voltage in the actual circuit follows that of the original model, it does so with the mentioned jumps, which necessitates another formulation of the membrane voltage's standard deviation, adapted from the original mean-field model in \cite{renart03}:
\begin{equation}
 \sigma_v = \sqrt{ ( f_\mathrm{syn}\cdot [\alpha\cdot (E_{syn}-\bar v)]^2 + f_\mathrm{mem}\cdot [\alpha\cdot (E_{syn}-\bar v)]^2 ) \cdot \tilde\tau_\mathrm{mem} }\;,
\end{equation}
where $f_\mathrm{syn}$ and $f_\mathrm{mem}$ are the switching frequencies for the synaptic and leak conductances, respectively.
They are calculated as follows:
\begin{equation}
 f_\mathrm{mem} = \frac{1}{\alpha\cdot \tau_\mathrm{mem}}\;,\;\;
 f_\mathrm{syn} = \frac{g_\mathrm{syn,total}}{g_\mathrm{mem}}\cdot f_\mathrm{mem}
\end{equation}
Please note that we use the same synapse type for both recurrent and background connections, so that there is only one switching circuit handling the synaptic input.

The modified variance formulation does not change the transfer curves significantly, but may explain some of the differences between measurements and original mean-field theory, as shown in the Results section.

%

%
%
%

%
%

\end{document}